\documentclass[11pt]{article}
\usepackage{theorem}
\newfont{\Bbb}{msbm10 scaled 1200}     
\newcommand{\mathbb}[1]{\mbox{\Bbb #1}}
\newcommand{\binom}[2]{{\left(\begin{array}{c} #1\\
 #2\end{array}\right)}}

\newcommand{\End}[2][]{\mathrm{End}_{#1}\left(#2\right)}
\newcommand{\Hom}[3][]{\mathrm{Hom}_{#1}\left(#2,#3\right)}
\newcommand{\Mat}[1]{\mathrm{Mat}(#1)}
{ \theorembodyfont{\normalfont}
\newtheorem{rem}{Remark} }

\def\p{\pi_x,\pi_y}
\def\Aff{\mathbb{A}}
\def\R{\mathbb{R}}
\def\C{\mathbb{C}}
\def\P{\mathbb{P}}
\def\Z{\mathbb{Z}}

\def\I{\mathcal{I}}
\def\N{\mathcal{N}}
\def\DM{\mathcal{D}_M}
\def\A{\mathcal{A}}

\def\D{\mathcal{D}}
\def\K{\mathrm{K}}
\def\koszul{\mathrm{Koszul}}

\def\e{\mathrm e}
\def\r{\mathrm r}
\def\l{\mathrm l}
\def\gr{\mathrm{gr}}
\def\gl{\mathrm{gl}}
\def\tot{\mathrm{tot}}

\def\im{\mathop\mathrm{im}}
\def\coker{\mathop{\mathrm{coker}}}
\def\min{\mathop\mathrm{min}}
\def\dim{\mathop\mathrm{dim}}
\def\span{\mathop\mathrm{span}}

\def\Ki{k[M_1\cup M_2]}
\def\Kii{k[M_1]\oplus k[M_2]}
\def\Kiii{k[M_1\cap M_2]}
\def\MV{Mayer-Vietoris}

\def\be{\begin{equation}}
\def\ee{\end{equation}}
\newcommand{\ba}[1][c]{\begin{array}{#1}}
\def\ea{\end{array}}

\begin{document}

\bigskip\bigskip
\begin{flushright}
\hfill ITEP-TH-80/01\\
\hfill ITP-2002-1E\\
\hfill hep-th/0201153\\
\end{flushright}
\bigskip\bigskip
\thispagestyle{empty}
\begin{center}
{\Large\bfseries On quantization of singular varieties\\[\smallskipamount]
 and applications to D-branes}\\
\bigskip\medskip
{\large Dmitry~Melnikov$^{a,\,b}$ and Alexander~Solovyov$^{c,\,d}$}\\
\bigskip\bigskip
\slshape
$^{a}$ITEP, Moscow, Russia;\\
e-mail: \textup{\ttfamily melnikov@gate.itep.ru}\\[\medskipamount]
$^{b}$Moscow State University, Department of Physics\\[\medskipamount]
$^{c}$Bogolyubov ITP, Kiev, Ukraine;\\
e-mail: \textup{\ttfamily solovyov@aep1.bitp.kiev.ua}\\[\medskipamount]
$^{d}$Kiev National University, Department of Physics\\
\end{center}

\bigskip\bigskip
\begin{abstract}
We calculate the ring of differential operators on some singular
affine varieties (intersecting stacks, a point on a singular curve
or an orbifold). Our results support the proposed connection of
the ring of differential operators with  geometry of D-branes in
(bosonic) string theory. In particular, the answer does know about
the resolution of singularities in accordance with the string
theory predictions.
\end{abstract}

\medskip


\newpage

\renewcommand{\thesubsection}{\arabic{subsection}}
\renewcommand{\thesubsubsection}
             {\arabic{subsection}.\arabic{subsubsection}}

\subsection{Introduction}

Merkulov~\cite{Me} proposed a construction for deformation
quantization of affine varieties. In
particular, he considered quantization of the
$n$-tuple point $x^n=0$ and has proved the quantum algebra of
functions on the associated phase space to be the matrix algebra
$\Mat{n}$. Quantization of $n$ coincident hyperplanes in
$\R^{N+1}$ gives the tensor product of the matrix algebra
$\Mat{n}$ and $N$ copies of the Heisenberg algebra related to
directions along the hyperplanes. This result reminds the
appearance of the non-Abelian degrees of freedom on the stack of
$n$ coincident D-branes \cite{Pol}, \cite{Wi3}.
Recently the relevance of this procedure
to D-brane physics (in particular, to boundary string field
theory \cite{Wi1}, \cite{Wi2}, \cite{Sh1}, \cite{Sh2})
was established quantitatively in~\cite{GS}.

The present note aims to apply this technique to some singular varieties
and to compare the result to that known from D-brane physics.

The paper is organized as follows. In the second section we
describe the procedure of quantization of affine varieties,
including singular ones.  The methods of algebraic geometry allow
us to perform the quantization. This is given in terms of the ring
of differential operators on a subvariety. We propose the explicit
description of this ring in terms of an arbitrary resolvent of the
ring of functions on the subvariety. According to the proposed
connection with the geometry of D-branes, the properties of the
ring of differential operators should capture, in
particular, the unbroken gauge symmetry in the open string sector.
In the third section we study several examples and compare
the results with the predictions of string theory. In  the case
of intersecting stacks  we find a good agreement with physics:
two subalgebrae responsible for the non-Abelian degrees of freedom
living on the worldvolume of each D-brane and non-local operators
(massive modes of the strings stretching between  branes). The
structure of the resulting algebra of differential operators
appears especially transparent in terms of  the \MV\ sequence. For a
point on a singular curve (``cusp'') the quantum algebra behaves
just as if the singularity was blown up (resolved). In the case of
the point on the $\C^2/\Z_m$ orbifold we find a perfect agreement
with the string theory picture: when the point approaches
the singularity, the dimension of the algebra increases $m^2$ times.
The same result is obtained using the blow-up.

Thus we found the complete agreement with bosonic string theory
picture. It would be interesting to generalize these
considerations to the case of superstrings. This will be done
elsewhere.

\subsection{Reduction of the ring of differential
operators onto a subvariety}
Algebraic geometry studies
geometric concepts using the technique of (commutative)
algebra. An important motivation
for such a study is application to reducible
or singular varieties (e.g.~\cite{MR}). The ring of polynomial
functions on an (affine algebraic) variety $M$ embedded into the
affine space $\Aff^N$ is given as the factor of the ring of
polynomials in $x^1,\ldots,x^N$ w.r.t.\ the ideal generated by the
set of equations defining $M$:
\be
\label{quot-func}
k[M]=k[x^1,\ldots,x^N]/(\phi_1,\ldots,\phi_n),
\ee
$\phi_\alpha(x)$ being some
polynomials and
$k$ being the
base field.
\be
\label{var-eq}
\left\{ \ba
\phi_1(x)=0,\\
\cdots\\
\phi_n(x)=0
\ea
\right.
\ee
are the equations of $M$.
Roughly speaking, $k[M]$ is the restriction of the
ring of polynomials $k[x^1,\ldots,x^N]$ onto $M$ having a
nontrivial kernel --- ideal $(\phi_1,\ldots,\phi_n)$. For
instance, let $X$ be the  $n$-tuple point given by the equation
$x^n=0$ on the affine line $\Aff^1$; then  for $n=1$
\be
k[M]\simeq k,
\ee
and for $n=2$
\be
\label{double point}
k[M]\simeq \left\{a+b\epsilon\,|\,a,b\in k,\:\epsilon^2=0\right\}.
\ee
So algebraically a point can be easily distinguished from a
double point which would never be possible in topological approach
to the set of solutions of (\ref{var-eq}).

We would like to define the ring of differential operators on
a subvariety. It is natural to use the construction of the ring of
functions on the subvariety in terms of some resolvent. Namely,
let ($K^\cdot,d_\K$) be a resolution of the structure sheaf $k[M]$
of the subvariety $M$ in the affine space $\Aff^N$ with $K^p\simeq
k[x]\otimes V_{n_p}$, $V_{n_p}\simeq k^{n_p}$, $\gr d_\K=-1$, and
$H^0(K^\cdot)\simeq k[M].$  We define differential operators
$\DM$ on the algebraic variety $M$ as the cohomologies
$H_\delta^{0,0}H_d$ of the double sequence
$\tilde{K}^{\cdot\cdot}$:
\be
\label{double complex} \tilde{K}^{p,q}\simeq \D\otimes
\Hom{V_{n_{q}}}{V_{n_{p}}}
\simeq V_{n_{p}}\otimes \D\otimes V_{n_{q}}^{\ast},
\ee
where $\D$ stands for the ring of differential
operators on $\Aff^N$. The horizontal differential acts according
to
\be
dm^{p,q}=d_\K\circ m^{p,q}\in\tilde{K}^{p-1,q},
\ee
and the vertical one
\be
\delta m^{p,q}=m^{p,q}\circ d_\K\in\tilde{K}^{p,q+1}
\ee
for $m^{p,q}\in K^{p,q}$.
Obviously, $d$ and $\delta$ commute.
The differential $d_\K$ of the
resolvent $K^\cdot$  is the multiplication by
a matrix whose entries are some functions,
and we regard those functions as the zeroth order
differential operators.
Thereby one needs to consider some matrix differential operators
as elements of $\tilde{K}^{\cdot\cdot}$
in order to define differential operators on a subvariety
(see (\ref{double complex}):
$\Hom{V_{n_{q}}}{V_{n_{p}}}\simeq\Mat{n_{p},n_{q}}$).

Note that thus defined differential operators on a subvariety
naturally act on the ring of functions on this subvariety. It is
obvious from the following general consideration. The double
complex just considered is a particular case of the general
construction (e.g.~\cite{RG}). Let $(K,d_\K)$ be a differential
module. These data determine the bicomplex
$(\tilde{K}=\End{K},d,\delta)$ with the two differentials
\be
\label{diff}
d:\End{K}\ni m\to d_\K\circ m, \quad
\delta:m\to m\circ d_\K
\ee
(note that $d_\K\in\End K$). In the ultimately
general case there exists the natural action of $H_\delta
H_d(\tilde{K})$ on $H(K)$ described as follows. Take a cohomology
class $[m]\in H_\delta H_d(\tilde{K});$ its representative $m$
satisfies
\be
\ba dm=0\iff d_\K\circ m=0,\\ \delta m=0\iff m\circ
d_\K=d_\K\circ\tilde{m} \;\mbox{for some $\tilde{m}$}
\ea
\ee
and is determined up to
\be
m\to m+d_\K\circ m_1+m_2\circ d_\K,\;\mbox{where}\; d_\K\circ
m_2=0.
\ee
Given $[f+d_\K g]\in H(K)$, the natural action
\be
[m][f]=[m f] \ee is defined unambiguously (that is why $\DM$
\emph{does} act correctly on $k[M]$). This fact is established by
the direct calculation. The key feature of the resolvent is its
acyclicity everywhere but in one degree of grading, so the same
should apply to the complex just considered: $(\End{K},d,\delta)$
also proves to be a resolvent with the nontrivial cohomologies
$H\simeq \End{K}$ (the injectivity of the just described canonical
homomorphism $\varphi:H(\End{K})\to\End{H(K)}$ is proved in
Appendix). Note that all the homomorphisms (endomorphisms)
are understood as those
over the base field $k$, $\Hom{\cdot}{\cdot}=\Hom[k]{\cdot}{\cdot}$.

In the above considerations we replace
$\End{k[x]}\to\D$ actually extracting a subclass $\D$ of all
endomorphisms $\End{k[x]}$ and call $H_\delta
H_d(\tilde{K}^{\cdot\cdot})$ differential operators on the
subvariety. It is completely in the spirit of algebraic geometry
--- recall the definition (\ref{quot-func}) of polynomial
functions on an affine variety.
Though the reduction procedure has the most natural
description in terms of all endomorphisms $\End{k[x]}$, i.e.\
all \emph{integral} operators, local differential operators are
of especial physical importance: these are local differential  operators
that describe massless modes (see further examples).

Now assume that constrains $\phi_\alpha$ are regular, i.e.\
$\phi_\alpha$ does not divide zero in
$k[x]/(\phi_1,\ldots,\phi_{\alpha-1})$. Consider the Koszul
resolvent $(K^\cdot,d_\K)$ relevant to this case~\cite{GH}:
\be
\begin{array}{ccccccccc}
\Lambda^{n}V\otimes k[x]& \stackrel{d_\koszul}{\longrightarrow} &
   \Lambda^{n-1}V\otimes k[x] & \stackrel{d_\koszul}{\longrightarrow} &
   \cdots & \stackrel{d_\koszul}{\longrightarrow} &
   k[x] & \longrightarrow & 0\;;
\end{array}
\ee
where $V=\span\{\e_1,\ldots,\e_n\}$ is the
$n$-dimensional vector space, $v=(\phi_1,\ldots,\phi_n)$; and the
differential $d_\koszul=i_v$ is given by the interior
differentiation, $i_v\e_\alpha=\phi_\alpha.$ The reduction
procedure is described by the double complex
$\tilde{K}^{\cdot\cdot}$:
\be
\begin{array}{ccccccccc}
\Lambda^{n}V\otimes\D\otimes\Lambda^{n}V^\ast &
       \stackrel{d}{\longrightarrow} &
       \Lambda^{n-1}V\otimes\D\otimes\Lambda^{n}V^\ast &
       \stackrel{d}{\longrightarrow} & \cdots &
       \stackrel{d}{\longrightarrow} & \D\otimes \Lambda^{n}V^\ast &
       \longrightarrow & 0 \\
\Big\downarrow\delta & & \Big\downarrow\delta & & & &
       \Big\downarrow\delta & & \\
\Lambda^{n}V\otimes\D\otimes\Lambda^{n-1}V^\ast &
       \stackrel{d}{\longrightarrow} &
       \Lambda^{n-1}V\otimes\D\otimes\Lambda^{n-1}V^\ast &
       \stackrel{d}{\longrightarrow} & \cdots &
       \stackrel{d}{\longrightarrow} & \D\otimes \Lambda^{n-1}V^\ast &
       \longrightarrow & 0 \\
\Big\downarrow\delta & & \Big\downarrow\delta & & & &
       \Big\downarrow\delta & & \\
\vdots\phantom{\delta} & & \vdots\phantom{\delta} & & &
       & \vdots\phantom{\delta} & & \\
\Big\downarrow\delta & & \Big\downarrow\delta & & & &
       \Big\downarrow\delta & & \\
\Lambda^{n}V\otimes\D & \stackrel{d}{\longrightarrow} &
       \Lambda^{n-1}V\otimes\D & \stackrel{d}{\longrightarrow} &
       \cdots & \stackrel{d}{\longrightarrow} &
       \D\phantom{\delta} & \longrightarrow & 0 \\
\Big\downarrow\phantom{\delta} & & \Big\downarrow\phantom{\delta} &
       & & &\Big\downarrow\phantom{\delta} & & \\
0\phantom{\delta} & & 0\phantom{\delta} & & & & 0\phantom{\delta} & &
\end{array}
\ee
The horizontal and vertical differentials $d$ and $\delta$ are
Koszul differentials, $d$ acting on $V$ and
$\delta$ on $V^\ast$
(to make notations for $\delta$ more convenient, we have
replaced $\Lambda^kV^\ast$ with its
Hodge dual $\Lambda^{n-k}V^\ast$).
$\D$ is a free right or left module over $k[x]$
(but it is not a free bimodule!);
so $H_d^{\cdot\cdot}(\tilde{K}^{\cdot\cdot})$
is concentrated in the last column, and
$H_\delta^{\cdot\cdot}(\tilde{K}^{\cdot\cdot})$ is concentrated in the
last row. These are the sufficient conditions that provide~\cite{BT}
\be
H_dH_{\delta}(\tilde{K}^{\cdot\cdot})\simeq H_D(A^\cdot)\simeq
H_{\delta}H_d(\tilde{K}^{\cdot\cdot}).
\ee
That is why $\DM$ does not depend on whether we work with the right or left
ideal (cf.~\cite{Me}).
The total complex is obtained via the contraction of the grading:
\be
\tot \tilde{K}^{\cdot\cdot}\simeq A^\cdot\simeq\bigoplus\limits_k{A^k},
\quad A^k=\bigoplus\limits_{p+q=k}{\tilde{K}^{p,q}}.
\ee
Its differential
\be
D=d+(-)^p\delta.
\ee

Let us stress that this definition of $\DM$ is equivalent
to that of~\cite{Me}:\footnote{In~\cite{Me} Merkulov used
deformation quantization and the Moyal $\star$-product. It leads
to integral operators in general, but if we restrict
ourselves to polynomials in $p,$ the Moyal algebra is nothing but
the algebra of differential operators $\D$ (see further
comments).}
\be
H_d^{0,n}\simeq \frac{\D}{\sum\limits_\alpha{\phi_\alpha \D}},
\ee
and the ring of differential operators on the subvariety is
constructed as
\be
\label{quot-op}
H_\delta^{0,n}H_d\simeq \DM\simeq \N/\I,
\ee
where the right ideal
\be
\I=\left\{\sum\limits_\alpha{\phi_\alpha\D}\right\}.
\ee
The associated normalizer
\be
\N=\left\{m\in \D\,|\,
m\phi_\alpha\subset\I\:\forall\alpha\right\}
\ee
is actually the
maximal subalgebra in which $\I$ is a two-sided ideal.
Introduction of normalizer makes the induced multiplication in
$\DM$ correctly defined. Such a prescription is precisely
the quantum analog of the Hamiltonian reduction~\cite{GS}.

Consider the codimension one example, i.e.\ a principal ideal $\I$
(e.g.\ $\I=(x^n)$). Contraction of the grading yields the total
complex $A^\cdot$ (physicists usually call it BRST):
\be
\ba
0\longrightarrow\D\stackrel{d_1}{\longrightarrow}\D\otimes
(k\oplus k)
\stackrel{d_2}{\longrightarrow}\D\longrightarrow 0,\\
d_1 f=(x^n f, f x^n),\quad d_2(g_1,g_2)=g_1 x^n-x^n g_2,\\
d_2\circ d_1=0.
\ea
\ee
What we need is $\ker d_2/\im d_1;$
\be
\ker d_2=\left\{ (g_1,g_2) \,|\, g_1x^n=x^n g_2\right\},
\ee
so a representative of $\ker d_2$ is uniquely determined by
$g_1\in\N_\r$ (of course, the subscript $_\r$ means ``right,''
whereas $_\l$ means ``left'').
\be
\im d_1=\left\{ (x^n f, fx^n) \,|\, f\in\D \right\},
\ee
and, finally,
\be
\ker d_2/\im d_1\simeq\DM.
\ee

One can express a representative of any (linear)
subspace in $\D$ modulo $\I$ as
\be
\label{gen-ans}
h(x,\partial)=h_0(\partial)+x h_1(\partial)+\cdots+x^{n-1}h_{n-1}(\partial)
\ee
for some $h_i=h_i(\partial)$. For $h\in\N$ this sum must also satisfy
the equation
(``$\equiv$'' means ``equal modulo $\I$,'' and $\I$ stands for the
ideal in the ring of functions or differential operators depending
on context)
\be
hx^n\equiv 0
\ee
or, equivalently (we denote $\partial=\pi$ for convenience),
\be
\label{h-eq}
\left\{
\sum\limits_{i=0}^{l}\binom{n}{l-i}
\frac{d^{n-l+i}h_i(\pi)}{d\pi^{n-l+i}}=0,
\right.
\quad l=0,\ldots,n-1.
\ee
We have used the commutation relation
\be
\label{power-lowering}
f(\pi)x^n=\sum\limits_{k=0}^n{\binom{n}{k}x^{n-k}\frac{d^k f}{d\pi^k}}\,.
\ee
The general solution is
\be
\label{gen-sol}
h_i(\partial) =\sum_{k=0}^{n-1} a_{i,k}\partial^k
+\sum_{k=0}^{i-1}\frac{1}{(i-k)!}
\left[\sum_{j=0}^{i-k-1}(-1)^{j+1}
\binom{i-k-1}{j} a_{k,n-i+k+j} \partial^{n+j}\right]
\ee
with $n^2$ arbitrary integration constants
$a_{i,k}$, $0\leq i,k\leq n-1$; thus $\dim\DM=n^2.$
The natural action of $\DM$ on $k[M]$ provides the
celebrated isomorphism $\DM\simeq \Mat{n}.$
We see that in the case of the
$n$-point local differential operators prove to be a sufficient
subclass of $\End{k[x]}$ to provide the complete matrix algebra,
i.e.\ all endomorphisms $\End{k[M]}.$

To complete the picture it remains to determine the
$\coker d_2\simeq\frac{\D}{\I_\l\oplus\I_\r}.$
The obvious relation
\be
\left(\sum\limits_{i=0}^{n-1}{x^i h_i}\right)x^n\equiv
\sum\limits_{l=0}^{n-1}\sum\limits_{i=0}^{l}
{\binom{n}{l-i}x^l\frac{d^{n-l+i}h_i}{d\pi^{n-l+i}}}
\ee
implies
\be
\frac{\D}{\I_\l\oplus\I_\r}=0. \ee

\subsection{Examples}
In this section we study some examples in order to compare the
result to that expected from physics. We find a perfect agreement.
In particular, it is clear how the non-Abelian degrees of
freedom appear on each of the intersecting stacks. The
quantization procedure also manages to find the zero
modes of the D-instanton. After that we examine the present
technique versus the resolution of singularities, and find that
the quantization knows about the resolution. This fact nicely
illustrates the connection of differential operators with string
theory.

\subsubsection{Intersecting stacks}
Consider the case of $\Aff^2$ and the unique constraint
(generator of the ideal)
$\phi=x^m y^n$. The calculation (see appendix B)
reveals that of importance are elements like (in this
subsection $h$'s are defined by (\ref{gen-ans}),(\ref{gen-sol}))
\be
\N\supset\span\{h(y,\partial_y)x^m f(x,\partial_x)\}.
\ee
They form a closed subalgebra $\A_y\subset\DM$:
\be
(h(y,\partial_y)x^m f(x,\partial_x))\,
(\tilde{h}(y,\partial_y)x^m\tilde{f}(x,\partial_x))
\equiv (h\tilde h)x^m(f x^m\tilde f).
\ee
This subalgebra is localized on the stack along the
$x$ axis: all the representatives vanish elsewhere.
There also exists a similar subalgebra
\be
\A_x=\span\{h(x,\partial_x)y^n f(y,\partial_y)\}.
\ee
The localization of these subalgebrae can be illustrated with the help of
the \MV\ sequence for $\DM$ constructed as follows.
The two ideals corresponding to the two stacks are
\be
\I_1=(x^m), \quad \I_2=(y^n).
\ee
The exact \MV\ sequence for $k[M]$
\be
\label{MVS}
\begin{array}{ccccccccc}
0 & \longrightarrow &
\Ki &
\stackrel{\mbox{embedding}}{\longrightarrow} &
\Kii &
\stackrel{\mbox{difference}}{\longrightarrow} &
\Kiii &
\longrightarrow & 0\\
 & & \parallel & & \parallel & & \parallel &  & \\
 & & \frac{k[x]}{\I_1\I_2}
 & & \frac{k[x]}{\I_1}\oplus\frac{k[x]}{\I_2}
 & & \frac{k[x]}{\I_1\oplus\I_2}  &  & \\
\end{array}
\ee
induces the acyclic double complex for $\End{k[M]}$
{\tiny
\be
\label{MVS for End}
\begin{array}{ccccccc}
\Hom{\Kiii}{\Ki} &
       \stackrel{d}{\longrightarrow} & \Hom{\Kiii}{\Kii} &
       \stackrel{d}{\longrightarrow} & \End{\Kiii} &
       \longrightarrow  & 0 \\
\big\downarrow\delta & & \big\downarrow\delta & &
       \big\downarrow\delta & & \\
\Hom{\Kii}{\Ki} &
       \stackrel{d}{\longrightarrow} & \End{\Kii} &
       \stackrel{d}{\longrightarrow} & \Hom{\Kii}{\Kiii} &
       \longrightarrow  & 0 \\
\big\downarrow\delta & & \big\downarrow\delta & &
       \big\downarrow\delta & & \\
\End{\Ki} &
       \stackrel{d}{\longrightarrow} & \Hom{\Ki}{\Kii} &
       \stackrel{d}{\longrightarrow} & \Hom{\Ki}{\Kiii} &
       \longrightarrow  & 0 \\
\big\downarrow\phantom{\delta} & & \big\downarrow\phantom{\delta} & &
       \big\downarrow\phantom{\delta} & & \\
0\phantom{\delta} & & 0\phantom{\delta} & & 0\phantom{\delta} & &
\end{array}
\ee}
with $d$ and $\delta$ defined by (\ref{diff}).
\begin{rem}
We must fulfil the condition \be \I_{1}\cap\I_{2}=\I_{1}\I_{2}
\ee for (\ref{MVS}) to be exact. That is why such a sequence is
of little use say for a double point with $\I_{1}=(x)=\I_{2}.$
\end{rem}
The short sequence (\ref{MVS}) provides the resolvent
\be
\begin{array}{ccccccc}
\Kii & \longrightarrow &
\Kiii & \longrightarrow & 0
\end{array}
\ee
with $H^{1}\simeq\Ki,$ $H^{2}\simeq0.$
The resolvent for $\End{\Ki}$ can be built
using the general recipe (\ref{diff}):
\be
\begin{array}{ccccc}
\Hom{\Kiii}{\Kii} &
       \stackrel{d}{\longrightarrow} & \End{\Kiii} &
       \longrightarrow  & 0 \\
\Big\downarrow\delta & &
       \Big\downarrow\delta & & \\
\End{\Kii} &
       \stackrel{d}{\longrightarrow} & \Hom{\Kii}{\Kiii} &
       \longrightarrow  & 0 \\
\Big\downarrow\phantom{\delta} & &
       \Big\downarrow\phantom{\delta} & & \\
0\phantom{\delta} & & 0\phantom{\delta} & &
\end{array}
\ee
Using (\ref{MVS for End}), one easily computes the
necessary cohomologies
\be
\ba
\End{\Ki}\simeq H^{2,1}_{\delta}H_{d}\simeq
\frac{d\Hom{\Kii}{\Ki}}{\delta d\Hom{\Kiii}{\Ki}}
\simeq
\frac{\Hom{\Kii}{\Ki}}{\delta\Hom{\Kiii}{\Ki}}\,.
\ea
\ee
In the case of the finite order operators
we have
\be
\End{\Ki}\simeq
\frac{\span\{(h_{x}h_{y}+\A_{x},\tilde{h}_{y}\tilde{h}_{x}+\A_{y})\}}
{\span\{(h_{x}h_{y},h_{x}h_{y})\}}\,;
\ee
and that is why we associate $\A_{x}$, $\A_{y}$ with
the two stacks and the finite-dimensional part
$\{h_{x}h_{y}\}$ with the intersection.

A very important property is
\be
\label{orthog}
\A_x\A_y=0=\A_y\A_x.
\ee
It allows to identify $\A_x$, $\A_y$ with the non-Abelian degrees
of freedom living on the worldvolume
of the corresponding stack of D-branes.
In the case of intersecting
stacks the following excitation
modes appear \cite{Wi3}, \cite{Dou1}, \cite{Dou2}:
\be
\left(
\begin{array}{cc}
\Mat{m} & \Psi \\
\Psi    & \Mat{n}
\end{array}
\right). \ee Orthogonal subalgebrae $\Mat{m}$, $\Mat{n}$ are
identified with $\A_x,$ $\A_y$. $\Psi$ corresponds to massive
strings stretching between the two stacks and may be interpreted
in terms of the non-local operators which appear if we omit the
locality condition in our construction.\footnote{Of course,
our results apply to the non-supersymmetric
(purely bosonic) theory.}

If the two stacks are intersecting at an arbitrary non-zero angle,
the result does not depend on its value. It also agrees well with
the answer known from physics: the structure of massless
modes does not depend on the non-zero angle between branes.

For an arbitrary number of stacks intersecting at a point
 the generator is given by $\phi=x^m y^n
\prod_{i=3}^{r}{\phi_i}$ with $\phi_1=x^m,$ $\phi_2=y^n,$
$\phi_i=(\alpha_i x+y)^{\beta_i}.$
The straightforward generalization is to consider
\be
\label{one-brane-algebra}
\A_x=\span\left\{\left(y^n\prod\limits_{i=3}^{r}{\phi_i}\right)
h(x,\partial_x)g(y,\partial_y)\right\}.
\ee
It is a closed subalgebra localized on the first stack.
This fact can be proved, for example, using (\ref{rot-th}).
Just as in the case of the two stacks,
the orthogonality relation
(\ref{orthog}) is satisfied for any pair of stacks.
Denote
\be
\label{two-branes-algebra}
\A_{ij}=\left\{
\phi_1\ldots\phi_{\hat{i}}\ldots\phi_{\hat{j}}\ldots\phi_r
H_{ij}(x,y,\partial)
\right\}\,,
\ee
where $H_{ij}$ satisfy
\be
H_{ij}\phi_i\phi_j=\phi_i\phi_j\tilde{H}_{ij}
\ee
for some $\tilde{H}_{ij}(x,y,\partial)\in\D.$ Obviously, $\A_i\subset\A_{ij}$
($\A_i$ was defined in (\ref{one-brane-algebra})).
Some orthogonality relations hold for $\A_{ij}$, e.g.\
\be
\A_{12}\A_{34}=\span\left\{
(\phi_3\phi_4\phi_5\ldots\phi_r H_{12})
(\phi_1\phi_2\phi_5\ldots\phi_r H_{34})
\right\}=0.
\ee
So one can associate $\A_{ij}$ with the pair of stacks $(i,j)$.

\subsubsection{A line with a double point}
The constraints are
\be
\left\{
\ba
xy=0,\\
y^2=0,
\ea
\right.
\ee
determining the ring of functions
\be
k[M]\simeq\left\{
f(x)+A\epsilon \,|\, \epsilon^2=0
\right\}.
\ee
The subring $\{f(x)\}$ is the ring of functions on
the affine line, and the nilpotent is responsible
for the D-instanton located at the origin.

Let us determine $\DM$.
Any operator modulo $\I$ can be brought to the form
\be
h=\sum_n{x^n f_n(\pi_x,\pi_y)}+yg(\pi_x,\pi_y).
\ee
For convenience we define $\pi_x=\partial_x$,
$\pi_y=\partial_y.$ Belonging of $h$ to $\N$ requires
\be
\left\{
\begin{array}{l}
\frac{\partial^2 f_n}{\partial\pi_y^2}=0, \quad n=0,1,\ldots,\\
\frac{\partial^2 g}{\partial\pi_y^2}+
      2\frac{\partial f_0}{\partial\pi_y}=0,\\
\frac{\partial f_{n-1}}{\partial\pi_y}+
      \frac{\partial^2 f_n}{\partial\pi_x\partial\pi_y}=0,
      \quad n=1,2,\ldots,\\
\frac{\partial^2 f_0}{\partial\pi_x\partial\pi_y}=0,\\
\frac{\partial^2g}{\partial\pi_x\partial\pi_y}+
      \frac{\partial f_0}{\partial \pi_x}=0.
\end{array}
\right.
\ee
The general solution in the class of finite order operators is
\be
h=f(x,\partial_x)(1-y\partial_y)+yg(\partial_x)+Cy\partial_y.
\ee

Obviously, $\{f(x,\partial_{x})(1-y\partial_{y})\}$ is the
subring of differential operators on the line, which can
be established through its action on $k[M]$:
\be
\ba
f(x,\partial_{x})(1-y\partial_{y})\,\phi(x)=
f(x,\partial_{x})\,\phi(x),\\
f(x,\partial_{x})(1-y\partial_{y})\,y=0.
\ea
\ee
It describes the original D-brane's degrees of freedom.
Analogously, $y\partial_{y}$ is related to the D-instanton,
\be
\ba
y\partial_{y}\,\phi(x)=0,\\
y\partial_{y}\,y=y,\\
(y\partial_{y})\circ(y\partial_{y})=y\partial_{y}.
\ea
\ee
At last, zero modes of the D-instanton
are the physical assignment of $\{yg(\partial_{x})\}$.

\subsubsection{A point on the cusp}
The cusp is defined by
\be \label{cusp} y^2-x^3=0. \ee
Additional
equation $x=a\neq0$ sets two different points on this curve. The
related algebra is the matrix algebra $\Mat{2},$ off-diagonal
elements being represented by non-local operators (shift
operators). When $a\to0$ these two points glue together to form a
double point with the ring of functions (\ref{double point}) and
the related algebra of differential operators $\DM\simeq\Mat{2}.$

Let us resolve the singularity. This goal is reached blowing up
the origin, i.e.\ saying that the good coordinate is
$s=\frac{y}{x}$ instead of $y$; then the equation of the curve
(\ref{cusp}) takes the form \be x^2 (x-s^2)=0. \ee The two
exceptional lines $x^2=0$ are discarded, and finally the curve
becomes the quadratic parabola $x=s^2$ after the resolution.
Combining this equation with the equation $x=a$, we again arrive
to the double point,
\be
\left\{
\ba[l]
s^2=0,\\
x=0,
\ea
\right.
\ee
in the $a\to 0$ limit.

It is not difficult to explain why it happens so.
As far as $a\neq 0$
the blow-up is a good diffeomorphism in
a neighbourhood of the point $(x=a,\,y=\sqrt{a^3}),$
and the smooth $a\to 0$ limit for $k[M]$ and $\DM$
(``smooth'' means the dimension, not the
structure constants!) is not surprising
for algebraic geometry ---
that is why one might expect the coincidence
of the two results for $\DM.$

The similar scenario takes place in the case of orbifolds
considered below.

\subsubsection{A point on the $\C^2/\Z_m$ orbifold}
Given the action of $\Z_m$ on $\C^2$ as
\be
\C^2\ni(u,v)\to(\epsilon u,\epsilon^{-1}v)\;
\mbox{with}\;\epsilon=\sqrt[m]{1}\,,
\ee
one has three invariants $x=u^m,$ $y=v^m,$ $z=uv$ satisfying
\be
z^m=xy;
\ee
this way the $\C^2/\Z_m$ orbifold is embedded into $\C^3$.
Here we dwell on the $\Z_2$ orbifold and
carry the more general case out into Appendix.

Consider first a quadruple-point
given by the set of equations
\be
\label{quadruple-point}
\left\{
\begin{array}{l}
x=1,\\
y^2=0,\\
z^2-xy=0.
\end{array}
\right.
\ee
The multiplicity is defined by splitting these coincident points,
e.g.\ via a small deformation of the r.h.s.
The element of the quotient ring of differential operators $\DM$
is represented as
\be
h\equiv\sum_{i=1}^{16}{C^i\e_i}
\ee
with the base vectors $\e_i$ given by
\be
\ba
\e_1=1, \quad
\e_2=y, \quad
\e_3=z, \quad
\e_4=yz, \quad
\e_5=zy\partial_z, \quad
\e_6=y\partial_z-zy\partial_z^{2}, \\
\e_7=\frac12yz\partial_z^{2}+yz\partial_y, \quad
\e_8=\frac16yz\partial_z^{3}+yz\partial_y\partial_z, \quad
\e_9=-y\partial_z^{2}+z\partial_z+\frac23 yz\partial_z^{3}, \\
\e_{10}=\frac16 y\partial_z^{3}+ y\partial_y\partial_z
       -\frac16 yz\partial_z^{4}-yz\partial_y\partial_z^{2}, \quad
\e_{11}=\partial_z+\frac23 y\partial_z^{3} -z\partial_z^{2}
       -\frac13 yz\partial_z^{4}, \\
\e_{12}=-\frac13 y\partial_z^{3}+
       \frac12 z\partial_z^{2}+ \frac{1}{12}yz\partial_z^{4}+
       z\partial_y-yz\partial_y^{2}-yz\partial_y\partial_z^{2}, \quad
\e_{13}=\frac12 y\partial_z^{2}+y\partial_y-\frac13 yz\partial_z^{3}, \\
\e_{14}=\frac12\partial_z^{2}+\frac{1}{12}y\partial_z^{4}+
       \partial_y-y\partial_y^{2}-y\partial_y\partial_z^{2}-
       \frac13 z\partial_z^{3}+\frac23 yz\partial_y\partial_z^{3}, \\
\e_{15}=-\frac16 y\partial_z^{4}-y\partial_y\partial_z^{2}+
       \frac16 z\partial_z^{3}+\frac{1}{12}yz\partial_z^{5}+
       z\partial_y\partial_z-yz\partial_y^{2}\partial_z+
       \frac13 yz\partial_y\partial_z^{3}, \\
\e_{16}=\frac16\partial_z^{3}+\frac{1}{12}y\partial_z^{5}+
       \partial_y\partial_z-y\partial_y^{2}\partial_z
       +\frac13 y\partial_y\partial_z^{3}
       -\frac16 z\partial_z^{4}-\frac{1}{36}yz\partial_z^{6}
       -z\partial_y\partial_z^{2}+yz\partial_y^{2}\partial_z^{2}.
\ea
\ee
The ring $\DM$ is isomorphic to $\Mat{4}$. To derive this representation,
the action of $\DM$ on $k[M]$ can be used again:
\be
\sum{C^i\e_i} \longleftrightarrow
\left(
\begin{array}{cccc}
C_1 & C_{14}     & C_{11}  & C_{16}               \\
C_2 & C_1+C_{13} & C_3+C_6 & C_{10}+C_{11}+C_{12} \\
C_3 & C_{12}     & C_1+C_9 & C_{14}+C_{15}        \\
C_4 & C_3+C_7    & C_2+C_5 & C_1+C_8+C_9+C_{13}
\end{array}
\right).
\ee
Thus the system (\ref{quadruple-point}) really defines four coincident
points; and if we are going to consider a single point, the appropriate
constraints are
\be
\left\{
\begin{array}{l}
y=a\neq0,\\
z=b,\\
z^2-xy=0.
\end{array}
\right.
\ee
In this case $\DM\simeq\C$ just as one expects for a single point.
What happens as the point approaches the singularity, i.e.\ $a\to0$?
The only way to set a point
sitting at the singularity is to put some
restrictions on both $x$ and $y,$ e.g.\
\be
\label{point at the singularity}
\left\{
\begin{array}{l}
x=0,\\
y=0,\\
z^2-xy=0.
\end{array}
\right.
\ee
Now one reads the resulting algebra as
\be
\DM\simeq\Mat{2}.
\ee
The geometric origin of this phenomenon is clear:
each singular point behaves as $m=2$ regular points glued together
causing the jump of $\dim\DM$ (the phenomenon is observed for
any $m$ --- see about this in appendix C).

The same result is obtained via the resolution technique.
The birational map is
\be
\label{blow-up}
\left\{
\ba[l]
y=vx,\\
z=wx.
\ea
\right.
\ee
Next we deform (\ref{point at the singularity}) substituting $x=a\to 0$
for $x=0.$ After the blow-up
\be
\left\{
\ba[l]
x=a\to 0,\\
y=0,\\
z^2=xy.
\ea
\right.
\to
\left\{
\ba[l]
x=a\to 0,\\
av=0,\\
w^2=v.
\ea
\right.
\to
\left\{
\ba[l]
x=a\to 0,\\
v=0,\\
w^2=0.
\ea
\right.
\ee
The latter is the proper equivalent set of equations
generating the same ideal and enjoying the smooth
$a\to 0$ limit that yields $\dim\DM=4.$

The ultimate question is why do we need to deform $x=0\to x=a$?
Such a behaviour  is typical for
blow-ups and is used for finding what points of the exceptional
line belong to our curve. To find whether a point belongs to
the curve, we move the point along the fiber of the tautological bundle
and check whether the moved point is far from the curve. (Recall
that geometrically the projective plane $\P^2$ that replaces the
origin of $\C^3$ after the blow-up is called the plane of
directions.)

\subsection*{Acknowledgements}
We are grateful to S.~Gukov for inspiring our work, and many
useful and encouraging discussions; and to A.~Gerasimov for his
extremely valuable discussions and remarks, in particular,
pointing out the importance of the locality condition and the
complex of endomorphisms, and carefully reading the manuscript. We
would like also to thank A.~Losev, A.~Levin, E.~Akhmedov,
A.~Chervov, S.~Oblezin, V.~Dolgushev, A.~Konechny, and D.~Belov
for many useful discussions and all ITEP group for their support
and stimulating. Support and discussions of A.~Morozov and
A.~Gorsky were crucial for the completion of the work. A.~S.\
would like to thank the Les Houches summer school and ITEP group,
where a part of the work was done, for the warm hospitality and
the very stimulating atmosphere. D.~M.\ was partially supported by
grants INTAS~\#~00-561, RFFI~\#~01-01-00549, and the grant of the
President of Russian Federation~\#~00-15-99296. A.~S.\ was
supported by grant INTAS~\#~99-590.

\appendix

\renewcommand{\thesubsection}{Appendix \Alph{subsection}}
\renewcommand{\thesubsubsection}
             {\Alph{subsection}\,\arabic{subsubsection}}

\subsection[\protect\hspace{1.7cm}
Proof of injectivity of the canonical homomorphism]
{Proof of injectivity of the canonical homomorphism}

We are going to verify that the kernel of the canonical
homomorphism $\varphi:H(\End{K})\to\End{H(K)}$
is trivial.
Let $[m]\in H_{\delta}H_{d}\tilde{K}$;
its representative is chosen up to
\be
m\to m^{\prime}=m+n\circ d_{\K}+d_{\K}\circ n^{\prime},
\;\mbox{where}\; d_{\K}\circ n=0.
\ee
The original differential module $K$ decomposes as
\be
K=\frac{K}{Z}\oplus Z,
\ee
where $Z=\ker d_{\K}.$ The reason for such a decomposition is that
in the case of interest the module $K$
is actually a vector space over the base field $k.$
Note that the image $\im m\subset Z$ since $d[m]=[0].$

The $n\circ d_{\K}$ term does not affect the action
of $m$ on $Z$; let us
investigate its action on $C\in \frac{K}{Z}:$
$mC+n\circ d_{\K}C\in Z.$ Picking up an appropriate
$n:B\to Z$ ($B=\im d_{\K}$ --- coboundaries),
we achieve
\be
(m+n\circ d_{\K})C=0\; \forall C\in\frac{K}{Z}
\ee
We have used the projectivity and the isomorphism
\be
\frac{K}{Z}\simeq d_{\K}\frac{K}{Z}\simeq B.
\ee
After such a choice of the representative $[m]\in\ker\varphi$ implies
$\im m\subset B$ requiring $m=d_{\K}\circ n^{\prime}$, so $[m]=[0]$ and
\be
\ker\varphi=[0].
\ee

\subsection[\protect\hspace{1.7cm}
Calculation of $\DM$ for intersecting stacks]
{Calculation of $\DM$ for intersecting stacks}
\subsubsection{Orthogonal stacks}
The right ideal and  normalizer
are defined as above:
\be
\I=x^m y^n \D,
\ee
\be
\N=\left\{f(x,y,\partial)\,|\,f x^m y^n=x^m y^n g\;
\mbox{for some}\; g\in\D\right\}.
\ee
Next we proceed to determine the moduli of the quotient $\DM\simeq\N/\I$.
In this case we can choose a representative of
$\N$ as
\be
f\equiv
\sum\limits_{k<m \mathrm{\;or\;}l<n}
{x^k y^l f_{kl}(\partial_{x},\partial_{y})}.
\ee
The condition $f\in\N$ after the repeated use
of the commutation relations (\ref{power-lowering}) yields
\be
0= \sum\limits_{l,k}
\sum\limits_{i=0}^m \sum\limits_{j=0}^n \binom{m}{i} \binom{n}{j}
x^{m+l-i} y^{n+k-j} \frac{\partial^{i+j}f_{lk}}{\partial{\pi_x^i}
\partial{\pi_y^j}},
\ee
which is equivalent to
\be
\label{inf-set}
\partial_{\pi_x}^m
\partial_{\pi_y}^n f_{\alpha\beta}=\\
-\underbrace{\sum\limits_{l=0}^{\min(m,\alpha)}
\sum\limits_{k=0}^{\min(n,\beta)}}_{k+l>0}
\binom{m}{l}\binom{n}{k}
\frac{\partial^{m+n-l-k}}{\partial\pi_x^{m-l}\partial\pi_y^{n-k}}
f_{\alpha-l,\beta-k},
\ee
where $\alpha<m$ or
$\beta<n$. So we obtain the following
recurrent relations:
\be
\ba
f_{\alpha\beta}= A_{\alpha\beta}^{(0)}(\pi_x)+
\cdots+A_{\alpha\beta}^{(n-1)}(\pi_x)\pi_y^{n-1}+
B_{\alpha\beta}^{(0)}(\pi_y)+\cdots+B_{\alpha\beta}^{(m-1)}
(\pi_y)\pi_x^{m-1}-\\
-\underbrace{\sum\limits_{l=0}^{\min(m,\alpha)}
\sum\limits_{k=0}^{\min(n,\beta)}}_{k+l>0} \binom{m}{l}\binom{n}{k}
\int\limits^{\pi_x}\!d\pi_x^{(1)} \ldots
\int\limits^{\pi_x^{(l-1)}}\!d\pi_x^{(l)}
\int\limits^{\pi_y}\!d\pi_y^{(1)} \ldots
\int\limits^{\pi_y^{(k-1)}}\!d\pi_y^{(k)}
f_{\alpha-l,\beta-k}(\pi_x^{(l)},\pi_y^{(k)}).
\ea
\ee
For example, in the case of $m=1$ and $n=1$
\be
\ba
f_{0,0}=A_0(\pi_x)+B_0(\pi_y),\\
\cdots\\
f_{k,0}=-\int\!d\pi_x\,f_{k-1,0}(\p)+A_k(\pi_x)+B_k(\pi_y),\\
\cdots
\ea
\ee
and the similar expressions are implemented for $f_{0,k}$.

The parametrization of the representative being fixed, a point of
the moduli space of $\DM$ proves to be the infinite number of some
functions; and it is not trivial to bring the algebra to a simpler
form. There arose such a complication because we had actually
dealt with some integral operators (differential operators of
infinite order) in this calculation. To understand this better,
consider two distinct points in $\Aff^1$,
\be
x^2-1=0.
\ee
Then the quotient algebra
\be
\DM=
\left\{
A+Bx+C(x-1)\e^{2\partial_x}+D(x+1)\e^{-2\partial_x}
\right\}
\simeq \gl(2)
\ee
is represented in terms of shift operators~\cite{GS}. Should we restrict
ourselves to the finite order operators, only the diagonal matrices would
survive. The finiteness of the order is a kind of the locality condition.

What are the finite order solutions of (\ref{inf-set})?
In the algebraic situation some maximal
$\alpha_0$ must exist, so that $f_{\alpha\beta}=0$ for $\alpha>\alpha_0$.
It allows us to set $\alpha=\alpha_0+m\geq m$ in (\ref{inf-set}). Only terms
with $l=m$ survive so
\be
\label{f-eq}
\sum\limits_{k=0}^{\beta}\binom{n}{k}
\frac{\partial^{n-k}f_{\alpha_0,\beta-k}}
{\partial \pi_y^{n-k}}=0, \quad
\beta=0,\ldots,n-1;
\ee
these equations coincide with (\ref{h-eq}) written
w.r.t.\ $\pi_y$ for $f_{\alpha_0,\beta}$. Now let $\alpha=\alpha_0+m-1$.
Taking into account (\ref{f-eq}) for $f_{\alpha_0,\beta}$, we obtain
the same equations for $f_{\alpha_0-1,\beta}$; then
for $f_{\alpha_0-2,\beta},\ldots$ Running over the same
procedure for $\beta=\beta_0,$ $\beta_0-1,$ $\ldots$, one establishes that
\be
\DM\simeq\span\left\{
h_y(x,\partial_x)y^n f_y(y,\partial_y)
+h_x(y,\partial_y)x^mf_x(x,\partial_x)+
\tilde{h}_y(x,\partial_x)\tilde{h}_x(y,\partial_y)
\right\},
\ee
where \mbox{$h$'s} satisfy (\ref{h-eq}) and are thereby given by
(\ref{gen-ans}), (\ref{gen-sol}).
Note that they are in the one-to-one correspondence with
the $m\times m$- (correspondingly $n\times n$-) matrices.
In the holomorphic case one can allow $f_{i}$ to be a power series
in $x^{i}$ ($i=x,y$).

\subsubsection{Two stacks intersecting
at an arbitrary angle}
\be
\phi=x^m(\alpha x+y)^n.
\ee
Performing the linear change of variables
\be
\left\{
\begin{array}{l}
x^{\prime}=x, \\
y^{\prime}=\alpha x+y,
\end{array}
\right.
\quad
\left\{
\begin{array}{l}
\partial_x^{\prime}=\partial_x-\alpha \partial_y, \\
\partial_y^{\prime}=\partial_y,
\end{array}
\right.
\ee
one finds that the algebra $\DM$ depends on no angular parameter $\alpha$.
In fact, in the present construction one needs no metric;
so all the transverse directions
to the stack are of equal importance. The isomorphism
$\DM(\alpha)\simeq\DM(\alpha=0)$ acts on the representatives as
\be
f^{(\alpha=0)}(x,y,\partial_x,\partial_y)\longleftrightarrow
f^{(\alpha)}(x,y,\partial_x,\partial_y)=
f^{(\alpha=0)}(x,\alpha x+y,\partial_x-\alpha \partial_y,\partial_y).
\ee
It will help us to derive the following
technical statement important in the future.
Consider again $m$ coincident
lines in $\Aff^2$ with $\phi=x^m.$ One can apply some speculations
concerning transversal coordinates: $\N$ modulo $\I$ is
\be
\N=\span\left\{
g(y,\partial_y)h(x,\partial_x)
\right\}+\I.
\ee
On making a change of coordinates
$x\to x,$ $y\to\alpha x+y,$ $\partial_x\to\partial_x-\alpha \partial_y,$
$\partial_y\to\partial_y$, $\N/\I$ becomes
\be
\N=\span\left\{
g(\alpha x+y,\partial_y)h(x,\partial_x-\alpha \partial_y)
\right\}+\I.
\ee
These two equations prove that for any $h(x,\partial_x),$ $g(y,\partial_y)$
there always exist some $\tilde h(x,\partial_x-\alpha \partial_y),$ $
\tilde g(\alpha x+y,\partial_y)$ such that
\be
\label{rot-th}
gh\equiv\tilde{g}\tilde{h}.
\ee

\subsection[\protect\hspace{1.7cm}More orbifolds]
{More orbifolds}
The set of equations defining the $n$-point
at the singularity of the $\Z_m$ orbifold is
\be
\label{n-point at the singularity}
\left\{
\ba[l]
x=0,\\
y^n=0,\\
z^m=xy
\ea
\right.
\iff
\left\{
\ba[l]
x=0,\\
y^n=0,\\
z^m=0.
\ea
\right.
\ee
The blow-up is again performed using (\ref{blow-up}),
and (\ref{n-point at the singularity}) takes the form
\be
\left\{
\ba[l]
x=a,\\
v^n a^n=0,\\
w^m a^{m}=0
\ea
\right.
\to
\left\{
\ba[l]
x=a\to 0,\\
v^n=0,\\
w^m=0.
\ea
\right.
\ee
We do not need to resolve the singularity completely: the unresolved
part of the singularity is located on the line at infinity
apart from the doings.

\end{document}